# Electron Spin Relaxation in Two Polymorphic Structures of GaN

Nam Lyong Kang

*Department of Nanomechatronics Engineering, Pusan National University, Miryang 627-706, Republic of Korea*

ABSTRACT

The relaxation process of electron spin in systems of electrons interacting with piezoelectric deformation phonons that are mediated through spin-orbit interactions was interpreted from a microscopic point of view using the formula for the electron spin relaxation times derived by a projection-reduction method. The electron spin relaxation times in two polymorphic structures of GaN were calculated. The piezoelectric material constant for the wurtzite structure obtained by a comparison with a previously reported experimental result was $P_{\text{pe}} = 1.5 \times 10^{29}$ eV/m. The temperature and magnetic field dependence of the relaxation times for both wurtzite and zinc-blende structures were similar, but the relaxation times in zinc-blende GaN were smaller and decreased more rapidly with increasing temperature and magnetic field than that in wurtzite GaN. This study also showed that the electron spin relaxation for wurtzite GaN at low density could be explained by the Elliot-Yafet process but not for zinc-blende GaN in the metallic regime.

Gallium nitride has attracted considerable attention because of its promising applications for optical devices, such as light-emitting diodes and blue lasers, and electronic devices for high power, high frequency and high temperature applications [1-6]. GaN exists in two allotropic forms, i.e. wurtzite and zinc-blende structures. The former is characterized by two lattice constants and is a thermodynamically stable phase. The latter, however, is characterized by a single lattice constant and is a metastable phase. The crystallographic symmetry of zinc-blende and wurtzite GaNs are also different. Therefore, the electron spin relaxation processes in both cases are expected to be different. Understanding the spin relaxation mechanisms in semiconductors is of great importance for the practical use of semiconductor spintronic devices because preserving the information injected into spin over a practical time scale is important for such devices.

Electron spin relaxations in semiconductors are governed by Yelliot-Yafet (EY) [7, 8] and Dyakonov-Perel (DP) [9] spin relaxation mechanisms. Very long electron spin relaxation times in highly $n$-doped bulk zinc-blende GaN [4, 6] have been reported, whereas short electron spin relaxation times were observed in wurtzite GaN [3, 5]. Therefore, it is important to determine the dominant spin relaxation mechanism under a range of conditions, such as temperature, external field and electron density, and to understand how the phonon and electron distribution functions are included because the condition-dependence of electron spin relaxation might be affected by the distribution functions. In piezoelectric materials, such as GaN, acoustic piezoelectric phonon scattering is the dominant scattering mechanism in the middle range of temperatures (50-150 K).

This paper reports that the formula for the electron spin relaxation time derived using Kang-Choi projection-reduction (KCPR) method can be interpreted from a fully microscopic point of view using a diagram, from which some intuition for the quantum dynamics of electrons in a solid can be obtained. The spin flipping and conserving processes can be explained in an organized manner because the formula includes the Planck and Fermi distribution functions in multiplicative forms (this is called the population criterion), which is relevant because phonons and electrons belong to different categories in a quantum-statistical classification. The formula is applied to a system of electrons interacting with piezoelectric deformation phonons mediated through the spin-orbit interaction. The piezoelectric material constants for wurtzite structure is determined by a comparison with the previous reported experimental results, and the temperature and magnetic field dependence of the electron spin relaxation times in two polymorphic structures of GaN were obtained.

When an electromagnetic wave of frequency $\omega$ is applied to a system of electrons interacting with piezoelectric deformation phonons mediated through spin-orbit interactions, the electron spin relaxation time $(T_1)$ for the EY process can be derived using the KCPR method and considering the Lorentzian approximation for weak electron-phonon interaction as follows [10]:

$$\frac{1}{T_1} = \frac{2\pi}{\hbar(f_{\alpha-} - f_{\alpha+})} \sum_{\gamma,q}$$

$$\times \big[\{(T_+(\alpha-,\gamma+) + T_-(\alpha-,\gamma+)\}\tilde{l}_z(\gamma+,\alpha+)$$

$$+\tilde{l}_+(\alpha-,\gamma\ +)\{T_+(\gamma+,\alpha\ +) + T_-(\gamma+,\alpha\ +)\}$$

$$+\tilde{l}_z(\alpha-,\gamma\ -)\{T_+(\gamma-,\alpha\ +) + T_-(\gamma-,\alpha\ +)\}$$

$$+\{T_+(\alpha-,\gamma-) + T_-(\alpha-,\gamma-)\}\tilde{l}_+(\gamma-,\alpha+)\big]$$

$$= A + B + C + D + E + F + G + H, \qquad (1)$$

where $f_{\alpha s}$ is the Fermi distribution function for an electron with energy $E_{\alpha s}$, where $s = +(-)$ for an up (down) spin, and $\tilde{l}_z(\alpha\pm,\beta\pm) \equiv 2|[l_z(q)]_{\alpha\beta}|^2$ and $\tilde{l}_+(\alpha-,\beta+) \equiv |[l^+(q)]_{\alpha\beta}|^2$ are the interaction coupling factors, where $[l_z(q)]_{\alpha\beta}$ and $[l^+(q)]_{\alpha\beta}$ are the matrix elements of $\mathbf{l}(q)$ given in Eq. (8), and $l^+(q) = l_x(q) + il_y(q)$. In Eq. (1), the transition factors, $T_+(\alpha s, \beta s')$ and $T_-(\alpha s, \beta s')$, are defined as

$$T_+(\alpha s, \beta s') \equiv \delta(\hbar\omega + E_{\alpha s} - E_{\beta s'} - \hbar\omega_q)P_+(\alpha s, \beta s') \qquad (2)$$

$$T_-(\alpha s, \beta s') \equiv \delta(\hbar\omega + E_{\alpha s} - E_{\beta s'} + \hbar\omega_q)P_-(\alpha s, \beta s'). \qquad (3)$$

Here $\delta(x)$ denotes the Dirac delta function, through which the energy conservation is maintained, and the population factors, $P_\pm(\alpha s, \beta s')$, are defined as follows:

$$P_+(\alpha s, \beta s') \equiv (1 + N_q)f_{\alpha s}(1 - f_{\beta s'}) - N_q f_{\beta s'}(1 - f_{\alpha s}) \qquad (4)$$

$$P_-(\alpha s, \beta s') \equiv N_q f_{\alpha s}(1 - f_{\beta s'}) - (1 + N_q)f_{\beta s'}(1 - f_{\alpha s}), \qquad (5)$$

where $N_q$ is the Planck distribution function for a phonon with energy, $\hbar\omega_q$. Eq. (4) corresponds to a transition from an $(\alpha, s)$ state to a $(\beta, s')$ state with phonon emission minus a transition from a $(\beta, s')$ state to an $(\alpha, s)$ state with phonon absorption and Eq. (5) is a reverse process. Therefore, Eqs. (4) and (5) satisfy the population criterion. The energy eigenvalue under a static magnetic field $B$ applied in the $z-$direction is given as $E_{\alpha s} = (n_\alpha + 1/2)\hbar\omega_c +$ $\hbar^2 k_{z\alpha}^2/2m_e + g\mu_B B s/2$, where $n_\alpha = 0,1,2,\cdots$, $\omega_c = eB/m_e$ is the cyclotron frequency, $m_e$ is the effective mass of an electron, $k_{z\alpha}$ is the $z$-component of the electron wave vector, $g$ is the electron g-factor, and $\mu_B$ is the Bohr magneton.

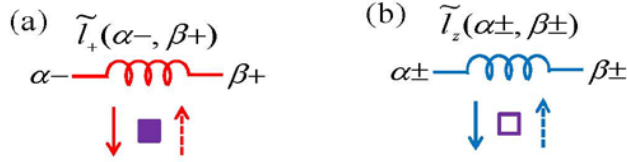

**FIG. 1.** Diagrammatic representation of the interaction factors, $\tilde{l}_+(\alpha-,\beta+)$ and $\tilde{l}_z(\alpha\pm,\beta\pm)$. The red spring emits (outward red solid arrow) or absorbs (inward red dotted arrow) a phonon denoted by a filled purple square, and the blue spring emits (outward blue solid arrow) or absorbs (inward blue dotted arrow) a phonon denoted by the empty purple square.

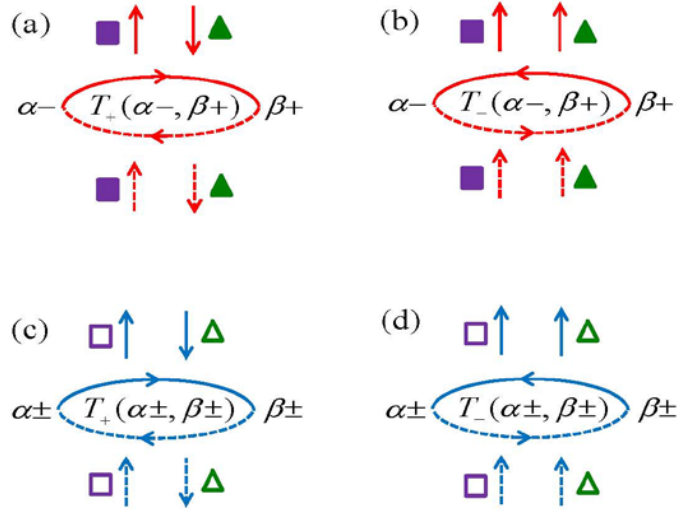

**FIG. 2.** Diagrammatic representation of the transition factors, $T_\pm(\alpha s, \beta s')$. $T_+(\alpha s, \beta s')$ and $T_-(\alpha s, \beta s')$ correspond to the clockwise and counterclockwise loops, respectively. A photon denoted by the filled green triangle is absorbed (emitted) during a forward (backward) process in the red loops, and a photon denoted by the empty green triangle is absorbed (emitted) during forward (backward) process in the blue loops. The upper and lower half circles correspond to the phonon emission and absorption processes, respectively.

For a diagrammatic interpretation of Eq. (1), the following are introduced [11]. $\tilde{l}_+(\alpha-,\beta+)$ and $\tilde{l}_z(\alpha\pm,\beta\pm)$ are represented by red and blue springs, respectively (Fig. 1). The red spring emits (outward red solid arrow) or absorbs (inward red dotted arrow) a phonon that is denoted by a filled purple square, and the blue spring emits (outward blue solid arrow) or absorbs (inward blue dotted arrow) a phonon that is denoted by an empty purple square. The filled and empty symbols are involved in spin flipping and conserving processes, respectively. Clockwise and counterclockwise loops are used to represent $T_+(\alpha s, \beta s')$ and $T_-(\alpha s, \beta s')$, respectively (Fig. 2). A photon denoted by a filled green triangle is absorbed (emitted) during the forward (backward) process in the red loops and a photon denoted by an empty green triangle is absorbed (emitted) during the forward (backward) process in the blue loops. The upper and lower half circles correspond to phonon emission and absorption processes, respectively.

Figure 3 presents a diagrammatic interpretation of the first (A) and third (C) terms in Eq. (1). The physical meaning of the first term (A) in Eq. (1) or Fig. 3 is as follows. The empty and filled black circles denote electrons with spin down and up, respectively. In a process from stages (1) to (2), $T_+(\alpha-,\gamma+)$ means that an electron with spin down transits from an initial spin down state, $\alpha-$, to an implicit spin up state, $\gamma+$, absorbing a photon and emitting a phonon to a lower spring and $\tilde{l}_z(\gamma+,\alpha+)$ means that the implicit state is coupled with a final spin up state, $\alpha+$, by a phonon with wave vector $q$ emitted from a lower spring. The $\gamma+$ and $\gamma-$ states are created by local fluctuations and are called the implicit states because they are contained in the relaxation time, not in the susceptibility tensor. A process from stages (2) to (3) is an inverse process that the electron transits from the final state to the implicit state absorbing a phonon emitted from the lower spring and emitting a photon. Combinations of the distribution functions in Eqs. (4) and (5) are different and the energies of the implicit states are determined by the energies of the final (or initial) state, photon and phonon. Therefore, a net transition is possible and the electron transition process forms a loop because phonon absorption and emission processes maintain a balance. The other processes in Eq. (1) can be interpreted from diagrams in a similar manner

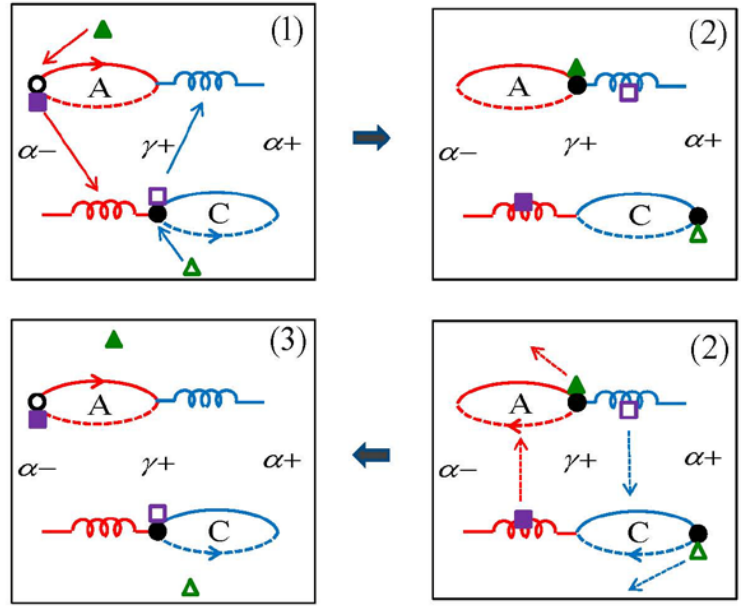

**FIG. 3.** Diagrammatic interpretation of processes A and C in Eq. (1). The black circle, green triangle and purple square denote the electron, photon and phonon. The filled (empty) triangles and squares are involved in the spin flipping (conserving) processes and the empty (filled) circles denote electrons with down (up) spins, respectively.

The electron spin relaxation times in two polymorphic structures of GaN are calculated using Eq. (1) for $n_\alpha = 0$ at the subband edge ($k_{z\alpha} = 0$) in the quantum limit. Only the cases $n_\gamma = 0$ and $n_\gamma = 1$ need to be considered in Eq. (1). A system of electrons interacting with piezoelectric deformation phonons through phonon-modulated spin-orbit interactions are considered, in which the interaction Hamiltonian is given as [8, 12]

$$V = \frac{\hbar}{4m_e^2 c^2}[\nabla V_{ep} \times (\mathbf{p} + e\mathbf{A})] \cdot \boldsymbol{\sigma}, \qquad (6)$$

where $c$ is the speed of light, $V_{ep}$ is the electron-phonon interaction potential, $\mathbf{p}$ is the momentum operator of an electron, $\mathbf{A}$ is the vector potential, and $\boldsymbol{\sigma}$ is the Pauli spin matrix. In a crystal whose lattice lacks inversion symmetry, such as GaN, the acoustic strain by pressure gives rise to a macroscopic electric field, which is assumed to be proportional to the derivative of the atomic displacement. Eq. (6) can then be expressed as

$$V = \sum_{\alpha,\beta} \sum_{s_\alpha,s_\beta} \sum_q [\mathbf{l}(q)]_{\alpha\beta} a^+_{\beta,s_\beta} a_{\alpha,s_\alpha} (b^+_{-q} + b_q) \cdot (\chi_{s_\alpha} \boldsymbol{\sigma} \chi_{s_\beta}). \tag{7}$$

Here, $a^+_{\alpha,s_\alpha}(a_{\alpha,s_\alpha})$ is the creation (annihilation) operator for an electron in the state, $|\alpha, s_\alpha>$, with a spin, $s_\alpha(= -$ or $+)$, $b^+_q(b_q)$ is the creation (annihilation) operator for a phonon in the state, $|q>$, with energy, $\hbar\omega_q$, $|q> \equiv |\mathbf{q}, l>$, $\mathbf{q}$ is the phonon wave vector, $l$ is the polarization index, $\chi_{s_\alpha}$ is the spinor, and

$$\mathbf{l}(q) = \frac{\hbar D_q}{4m_e^2 c^2} [\nabla e^{iq \cdot r} \times (\mathbf{p} + e\mathbf{A})], \tag{8}$$

where $D_q = P_{pe} q\sqrt{\hbar/(2\rho_m \Omega_0 \omega_q)}/(q^2 + q_d^2)$, $\rho_m$ is the mass density, $\Omega_0$ is the volume of the system, and $q_d = \sqrt{n_e e^2/\kappa \varepsilon_0 k_B T}$ is the reciprocal of the Debye screening length, where $\kappa$ is the static dielectric constant and $n_e$ is the density of electrons. In this paper, the proportional constant (piezoelectric material constant), $P_{pe}$, is used as a fitting parameter. This constant affects only the magnitude of the spin relaxation time, i.e., it does not affect the temperature dependence of the spin relaxation time because it is not contained in the distribution functions for electrons and phonons.

$P_{pe} = 1.5 \times 10^{29}$ eV/m for wurtzite GaN was obtained from Fig. 4 by fitting the present theoretical result to the experimental result reported by Ishiguro et al. [5] for $n_e = 1 \times 10^{20}$ m$^{-3}$. This theory was fitted to the experimental data at high temperatures (above 40 K) because the piezoelectric phonon scattering is dominant at these temperatures. The discrepancy at low temperatures (below 40 K) may be corrected if the impurity and acoustic deformation phonon scatterings or the DP mechanism are considered. The relaxation times increase with increasing electron density because the effect of phonon scattering decreases with increasing screening effect as the electron density is increased. Note that the relaxation time is an inverse of the scattering effect (or relaxation rate). The effect of phonon scattering increases with increasing number of phonons as the temperature is increased. Therefore, the relaxation times decrease with increasing temperature. Moreover, this decrease occurs sharply at high electron densities because the reciprocal of the Debye screening length decreases more rapidly with increasing temperature as the electron density is increased. The temperature dependence of the relaxation time for $n_e = 10 \times 10^{20}$ m$^{-3}$, $T^{-1.31}$, is similar to the result reported by Kuroda et al., $T^{-1.4}$ [3].

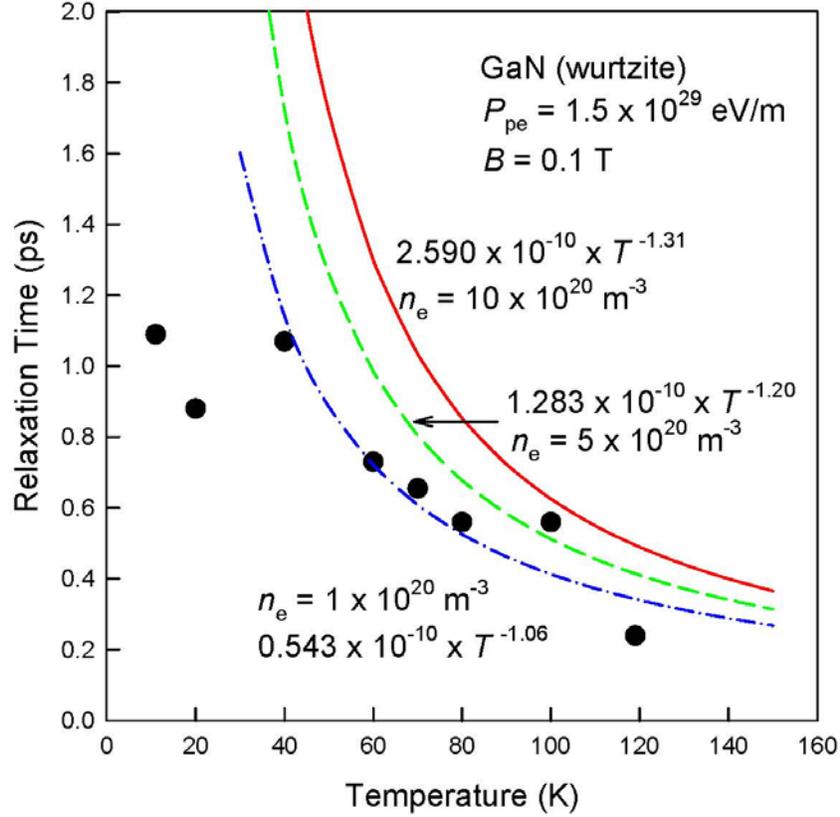

**FIG. 4.** Temperature dependence of the electron spin relaxation times in wurtzite GaN for $P_{\text{pe}} = 1.5 \times 10^{29}$ eV/m and various electron densities at $B = 0.1$ T. The black circles are the results reported by Ishiguro et al. [5].

Figure 5 presents the relaxation times for zinc-blende GaN, where the same piezoelectric material constant was used because there was no comparative experimental data. As mentioned above, the constant does not affect the temperature dependence of the relaxation time. Zeeman splitting in zinc-blende GaN is larger than that in wurtzite GaN because the effective mass of an electron in wurtzite GaN is larger than that in zinc-blende GaN. Therefore, the relaxation time in zinc-blende GaN is smaller and decreases more sharply with increasing temperature than that in wurtzite GaN because the electrons in zinc-blende GaN are scattered by phonons with high energies, and the number of phonons with higher energies increase more rapidly with increasing temperature than those with lower energies.

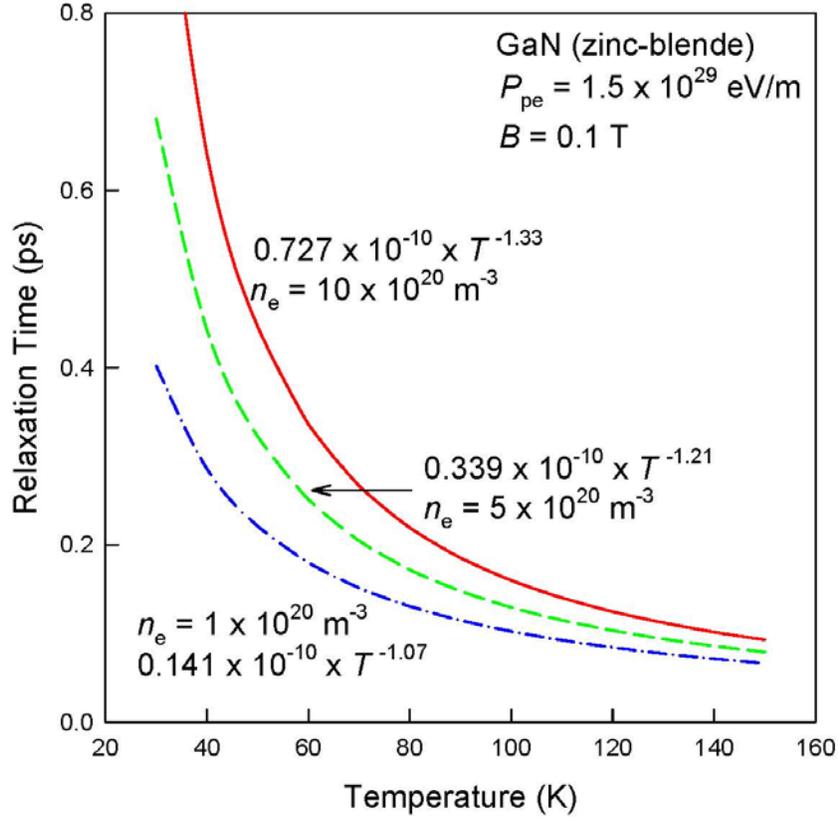

**FIG. 5.** Temperature dependence of the electron spin relaxation times in zinc-blende GaN for $P_{\mathrm{pe}} = 1.5 \times 10^{29}$ eV/m and various electron densities at $B = 0.1$ T.

Figures 6 and 7 present the magnetic field dependence of the relaxation times in the wurtzite and zinc-blende GaNs at 100 K. The relaxation times decrease with increasing magnetic field because electrons are scattered by phonons with high energies as the spacing between the energy levels increase with increasing magnetic field. The relaxation time in zinc-blende GaN decreases more rapidly than that in wurtzite GaN with increasing magnetic field for the same reason shown in Figs. 4 and 5. Table I lists the temperature and magnetic field dependence of the relaxation times in the wurtzite and zinc-blende GaNs.

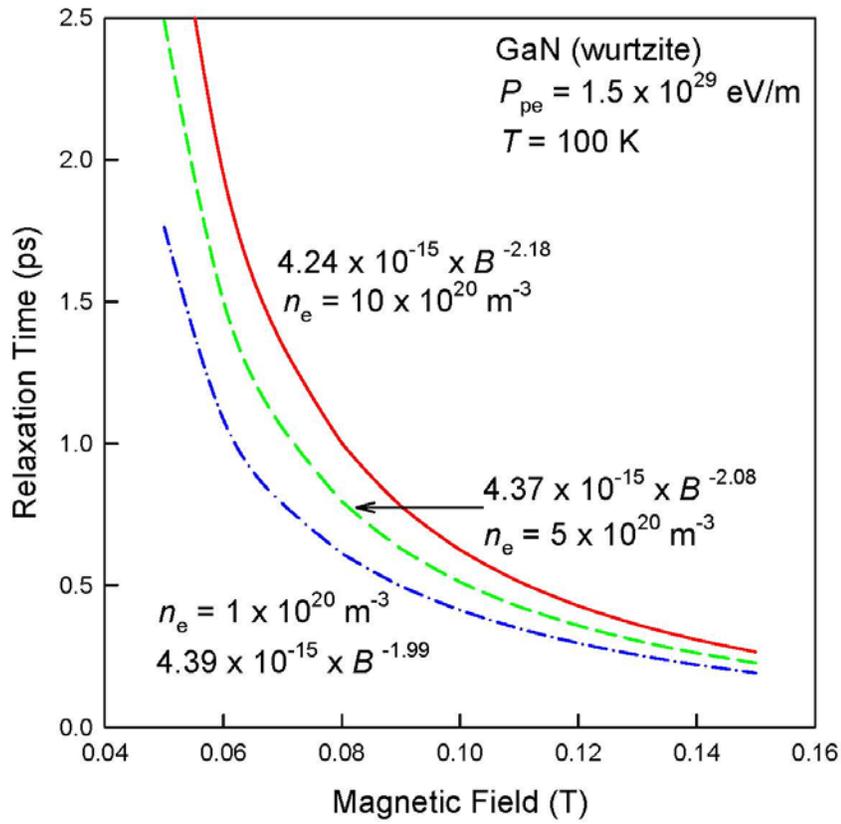

**FIG. 6.** Magnetic field dependence of the electron spin relaxation times in wurtzite GaN for $P_{\text{pe}} = 1.5 \times 10^{29}$ eV/m and various electron densities at $T = 100$ K.

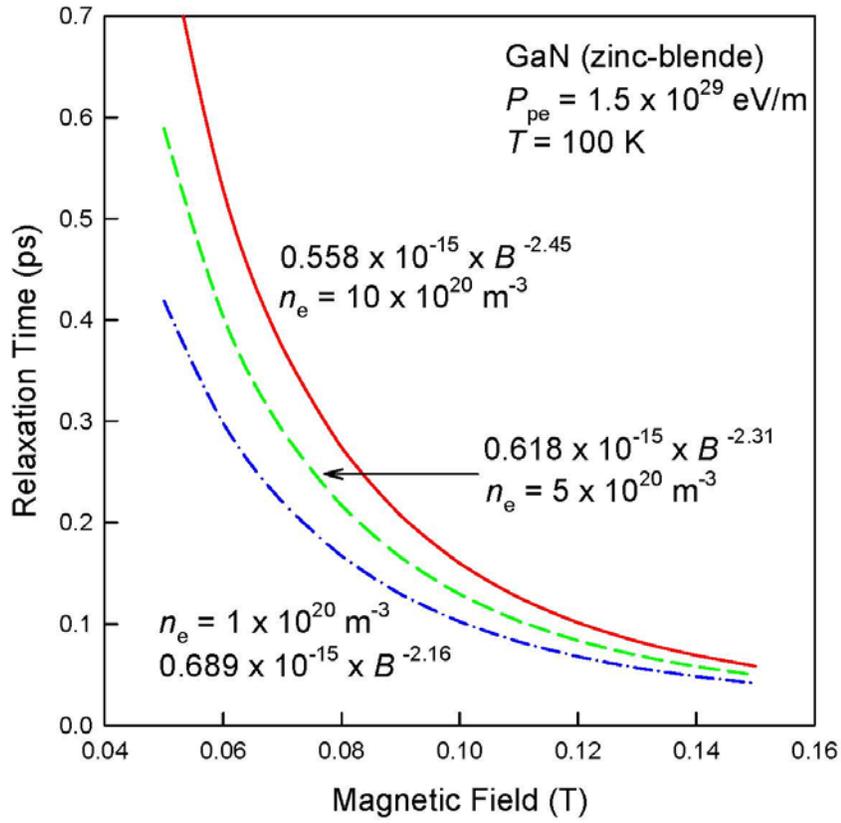

**FIG. 7.** Magnetic field dependence of the electron spin relaxation times in zinc-blende GaN for $P_{\text{pe}} = 1.5 \times 10^{29}$ eV/m and various electron densities at $T = 100$ K.

**TABLE I:** Temperature and magnetic field dependence of the spin relaxation times in two polymorphic structures of GaN. The temperature dependence and the magnetic field dependence were obtained at $B = 0.1$ T and $T = 100$ K, respectively [$T$: temperature, $B$: magnetic field, $T_1$: relaxation time, $n_e$: electron density].

| $n_e$ [m$^{-3}$] | wurtzite | zinc-blende |
|---|---|---|
| $1 \times 10^{20}$ | $T_1 \propto T^{-1.06}$ | $T_1 \propto T^{-1.07}$ |
| | $T_1 \propto B^{-1.99}$ | $T_1 \propto B^{-2.16}$ |
| $5 \times 10^{20}$ | $T_1 \propto T^{-1.20}$ | $T_1 \propto T^{-1.21}$ |
| | $T_1 \propto B^{-2.08}$ | $T_1 \propto B^{-2.31}$ |
| $10 \times 10^{20}$ | $T_1 \propto T^{-1.31}$ | $T_1 \propto T^{-1.33}$ |
| | $T_1 \propto B^{-2.18}$ | $T_1 \propto B^{-2.45}$ |

In conclusion, this paper showed that the formula for the electron spin relaxation time derived using the Kang-Choi projection-reduction method for a system of electrons interacting with piezoelectric phonons mediated through spin-orbit interaction can explain the electron spin flipping and conserving processes from a macroscopic point of view by diagrams. The piezoelectric material constant ($P_{\text{pe}}$) for wurtzite GaN obtained by fitting the present theoretical result to the previous reported experimental data for $n_e = 1 \times 10^{20}$ m$^{-3}$ was $P_{\text{pe}} = 1.5 \times 10^{29}$ eV/m. The respective temperature and magnetic field dependence of the electron spin relaxation times for $n_e = 1 \times 10^{20}$ m$^{-3}$ were $T_1 \propto T^{-1.06}$ and $T_1 \propto B^{-1.99}$ in wurtzite GaN, and $T_1 \propto T^{-1.07}$ and $T_1 \propto B^{-2.16}$ in zinc-blende GaN. The electron spin relaxation times for zinc-blend GaN are smaller and decrease more rapidly with increasing temperature and magnetic field than those for wurtzite GaN. Although this paper attributed these to the effective mass of an electron and the number of phonons, $P_{\text{pe}}$ may be another reason. Nevertheless, this was not confirmed because the experimental data for zinc-blende GaN was unavailable, and thus $P_{\text{pe}}$ for this material could not be obtained. The long electron spin lifetimes in highly doped cubic GaN by Buß et al. [6] could not be explained by the present formula including the Elliot-Yafet mechanism. Therefore, the relaxation of electron spin in metallic regime cannot be caused by the Elliot-Yafet mechanism. This phenomenon may be explained by the D'yakonov-Perel mechanism or electron-electron interaction, which will be studied in the future using the present projection- reduction method.

**References**

[1] M. Fanciulli, T. Lei, and T. D. Moustakas, Phys. Rev. B **48**, 15144 (1993-II).

[2] S. Nakamura, T. Mukai, and M. Senoh, Appl. Phys. Lett. **64**, 1687 (1994).

[3] T. Kuroda, T. Yabushita, T. Kosuge, A. Tackeuchi, K. Taniguchi, T. Chinone, and N. Horio, Appl. Phys. Lett. **85**, 3116 (2004).

[4] A. Tackeuchi, H. Otake, Y. Ogawa, F. Takano, and H. Akinaga, Appl. Phys. Lett. **88**, 162114 (2006).

[5] T. Ishiguro, Y. Toda, and S. Adachi, Appl. Phys. Lett. **90**, 011904 (2007).

[6] J. H. Buß, J. Rudolph, T. Schupp, D. J. As, K. Lischka, and D. Hägele, Appl. Phys. Lett. **97**, 062101 (2010).

[7] R. J. Elliott, Phys. Rev. **96**, 266 (1954).

[8] Y. Yafet, in *Solid State Physics*, ed. F. Seitz and D. Turnbull (Academic, New York, 1963), Vol. 14, pp. 1-98.

[9] M. I. D'yakonov and V.I. Perel', Zh. Eksp. Teor. Fiz. **60**, 1954 (1971).

[10] N. L. Kang and S. D. Choi, Chin. Phys. B **22**, 087102 (2014).

[11] N. L. Kang and S. D. Choi, Jpn. J. Appl. Phys. **53**, in press (2014).

[12] O. D. Restrepo and W. Windl, Phys. Rev. Lett. **109**, 166604 (2012).